\documentclass[a4paper,english,11pt]{article}

\usepackage[text={174mm,258mm},%
  papersize={210mm,297mm},%
columnsep=12pt,%
headsep=21pt,%
centering]{geometry}

\usepackage{amsmath,amssymb,bm,xcolor} %

\usepackage{url}

\usepackage{booktabs} %
\usepackage{xtab} 
\usepackage{dcolumn} 
\usepackage{array} 
\usepackage{graphicx}

\usepackage{authblk}

\usepackage[backend=biber,style=nature,autocite=superscript,intitle=true,url=false,doi=true,isbn=false,date=year,sortcites=false,eprint=false]{biblatex} %
\usepackage{hyperref}
\hypersetup{colorlinks=true, citecolor=blue, linkcolor=red, urlcolor=blue }

\usepackage{titlesec}
\newcommand{\addperiod}[1]{#1.}
\titleformat{\subsubsection}[runin] {\normalfont\normalsize\bfseries}{\thesubsubsection}{1em}{\addperiod}

\title{Learning about the liveability of cities from young migrants using the combinatorial Hodge theory approach}
\author[1]{Takaaki Aoki$^*$}
\author[2]{Kohei Nagamachi}
\author[2]{Tetsuya Shimane}
\affil[1]{Faculty of Data Science, Shiga University, Hikone 522-8522, Japan.}
\affil[2]{Graduate School of Management, Kagawa University, Takamatsu 760-8523, Japan.}
\date{Corresponding author: takaaki.aoki.work@gmail.com}

\begin{document}
\maketitle
\begin{abstract}
In declining and ageing societies, local communities face the `risk of eventual extinction.'
In Japan, a population equivalent to that of an entire city is lost every year, representing one of the most severe cases of population decline.
Thus, attracting young people has become a policy priority for many local municipalities in Japan, prompting the implementation of numerous initiatives to improve the liveability of affected cities.
However, what exactly is a liveable city? 
To determine this, a concrete measure of liveability is required to serve as a key performance indicator (KPI) for local governments to adopt.
In this study, we propose empirical liveability based on people's votes with their feet, following Tiebout's argument \autocite{tiebout1956pure}, and derive that such liveability can be quantified using the `potential' in the combinatorial Hodge theory, directly calculated from migration data only.
As a case study, we measure the empirical liveability of municipalities in Japan for specific populations --- families with small children and women of reproductive age.
Then, using the empirical liveability as dependent variables, we perform a regression analysis to identify factors related to liveability.
This method is applicable to various datasets on migration, categorized by ethnicity, education, skill level, income and other attributes, and provides valuable statistics for urban planning and policymaking.
\end{abstract} 

\section*{Keywords}
Hodge decomposition, internal migration, liveable cities, mobility, potential

\section{Introduction}
Several troubling statistics reflect the challenges faced by Japanese society today. 
Specifically, the population of Japanese nationals has been decreasing for the past 15 years, falling by more than 900,000 yearly, according to the most recent data from 1 January, 2025 \autocite{jpcensus}.
This yearly loss is comparable to the population of a city.
The Population Strategy Council, a non-governmental advisory group, reported that
744 of 1729 local municipalities nationwide have been identified as being `at risk of eventual extinction' \autocite{masudareport}.
This issue is particularly linked to the fact that the population of young women aged 20 to 39 is projected to decrease by half by 2050.
This is a widely recognized and significant issue in Japanese society as well as
a leading example of advanced population decline and demographic aging among developed nations.
Many countries in the OECD also face declining and ageing populations, with some regions projected to lose 20\% or more of their population by 2050 \autocite{ocedreport}.
Without gaining young people, the continuity of local Japanese governments and municipalities is under threat; therefore, they are implementing child-rearing support and migration policies to improve liveability.
However, what exactly is a liveable city? 
Liveability is complex and related to many economic and social factors of cities.
Beginning with seminal works on internal migration and urban studies, researchers have attributed migration to the regional differences in each city's various abilities,
which are related to  income and job opportunities \autocite{hicks1963theory,schultz1945agriculture,harris1970migration}, amenities \autocite{blomquist1988new,diamond2016determinants},
and place utility as a composite of these factors \autocite{wolpert1965behavioral},
leading to the expectation that people are commonly motivated to move from worse to better locations, beyond preferential variations among individuals (see Greenwood\autocite{greenwood1985human}, Aoki \& Inamura \autocite{AokiInamura1997}, and Cushing \& Poot \autocite{cushing2004crossing} for extensive reviews of relevant works).
An ambiguous goal of becoming a liveable city makes it hard to accomplish the intended outcome.
Therefore, a concrete key performance indicator (KPI) for liveability is needed for local government policymakers.
This motivates our central research question: `What KPI should be used to guide the implementations of policies aimed at enhancing liveability?'

To answer this question, we follow Tiebout's argument of `voting with one's feet', to uncover a concrete measure of liveability which explains migration flows among places.
Internal migration flows are the results of people's assessments of their living places.
Migration, leaving one place and moving to another, often incurs considerable pecuniary and non-pecuniary costs.
For this, people critically evaluate which cities would be preferable as their living places of residence.
Migration data are typically represented by an origin--destination (OD) matrix $M$, whose element $M_{ij}$ indicates the number of migrants from origin $i$ to destination $j$.
Therefore, the OD matrix indicates `people's votes with their feet' \autocite{tiebout1956pure,douglas1997estimating}, demonstrating that the destinations are more liveable than the origins. 
These movement data provide more substantial evidence of a location's liveability than do perceptual evaluations or stated interests collected in surveys.
This empirical measure of liveability thus provides a valuable metric for policymakers.

However, a technical problem remains: how should liveability from migration data be quantified?
Migration flows from origins to destinations only represent the comparison between the locations, not an indicator for each location: the liveability of each city is hidden in the data.
Therefore, it is important to extract a location-level score $s_i$, whose regional difference $s_j - s_i$ explains the network-level variable $M_{ij}$ between locations, linking the locations of migration.

Thus, we introduce the `potential' of migration flow.
Scalar potential, or potential, is a well-known mathematical concept adopted in various scientific fields and provides an intuitive representation of the hidden abilities from which flow is generated.
Specifically, we introduce the mathematical framework of the combinatorial Hodge theory \autocite{Jiang2011b, aoki2022urban,aoki2023identifying} into the analysis of migration flow.
A given flow on a network is uniquely decomposed into a gradient component described by the difference in potentials and the other circular components. As described later, the potential of the unique decomposition extracts the regional differences that reproduce the imbalanced acyclic component of the migration flow and quantifies the liveability of each location.

We demonstrate how to use this extraction method in real-world situations. As a case study, we evaluate potential as the empirical liveability of Japanese municipalities for two specific populations related to sustainability: families with small children (FwSC) and women of reproductive age (WoRA), aiming to contribute to local government policymaking.
In Japan, a municipality is the lowest level of government, with its own elected mayors responsible for local administration and services, and a city is a type of municipality on equal footing with towns and villages but legally distinguished by its larger population and urban characteristics (see `Methods' Section for details).
Then, using potentials as dependent variables, we perform a regression analysis to identify and clarify the economic and social factors that are relevant to liveability, including public services and institutions by local governments, such as schools, hospitals and libraries.

This study makes two key contributions to the existing literature.
First, empirical liveability is quantified based on people's votes with their feet. The metric is simply given as location-level descriptive statistic, directly calculated from the observed migration data without additional information on cities.
This contrasts with the micro-founded, elaborate modelling approach used in urban and regional economics literature, in which individuals' rational decisions are structurally modelled and closely integrated with several relevant datasets on cities \autocite{cushing2004crossing,biagi2018theoretical}.
Migration flows are aggregates of individuals' rational decisions \autocite{sjaastad1962costs}.
Individuals choose their residential locations by considering interregional utility differentials, which depend on regional differences in economic factors such as 
wages and job opportunities \autocite{harris1970migration,greenwood1989jobs}
and cost of living, as well as non-economic factors, including natural amenities \autocite{graves1979lifecycle,graves1980migration,rappaport2007moving},
consumption amenities, such as restaurants and museums \autocite{glaeser2001consumer}
and local public goods/bads such as public schools and crime \autocite{diamond2016determinants}.
In addition to this typical setting, individual heterogeneity, such as age \autocite{sjaastad1962costs}, human capital \autocite{faggian2009human,brown2012human}, and preference \autocite{douglas1993voting,nakajima2011estimating} can also be considered.
The structural estimation approach allows us to evaluate the effects of these factors on the mechanisms of migration flows \autocite{crozet2004migrants,pons2007testing} (There is also the growing literature of `quantitative spatial economics' \autocite{redding2017quantitative},
in which more elaborate general equilibrium models were developed to conduct a quantitative analysis \autocite{caliendo2019trade}, for example, considering migration as a households' dynamic decision problem to assess the impacts of the `China trade shock' on US local labour markets, which differ in industrial structure and trade frictions).
Second, the introduced mathematical framework presents a theoretical viewpoint on existing measures on migration.
Net migration, a traditional measure of migration which is the difference between in-migration and out-migration, is proved to be the potential in the combinatorial Hodge theory in the specific condition that distance factor of migration is ignored as discussed later.
Another metric is the interregional utility differentials estimated from migration data. 
Based on Samuelson's revealed preference approach\autocite{samuelson1948consumption} and Tiebout's voting with one's feet approach\autocite{tiebout1956pure}, Douglas \& Wall\autocite{douglas1993voting} and Douglas\autocite{douglas1997estimating} proposed a ranking algorithm for comparison among regions in terms of utility (or the `standard of living' in their term) and applied it to Canadian and US data, respectively.
Later, they developed a regression analysis to estimate utility differentials and applied it to Canadian and UK data \autocite{douglas2000measuring,wall2001voting}.
Nakajima \& Tabuchi \autocite{nakajima2011estimating} further developed another utility estimation formula improving the previous studies based on some stylised facts suggesting the type of migration costs.
We contribute to the literature by showing that the utility differentials obtained by the ordinary least squares (OLS) estimation in the previous works\autocite{douglas1997estimating,nakajima2011estimating} have an analytical and intuitive closed-form expression.
We also prove that the utility is a potential with globally consistent rankings and one that matches the gradient component of a unique decomposition of migration flows on a network under a specific migration flow indicator as shown in the `Methods' Section.

The remainder of this paper is organised as follows.
First, we describe the theoretical results, elucidating how to derive consistent information on empirical liveability from people's migration data.
Next, we explain how the proposed method relates to the existing measures on migration.
Then, we demonstrate how the proposed method can be used in a practical situation.
For this, we evaluated the empirical liveability of Japanese municipalities, particularly for FwSC and WoRA. %
Then, we report on the regression analysis to determine which economic and social factors are essential for liveability. 
Finally, we summarise and discuss the findings.

\section{Results}
\subsection{How can a consistent measure from people's migration flow be derived?}
\label{sec:theoretical_results}

\begin{figure}
  \includegraphics[width=\linewidth]{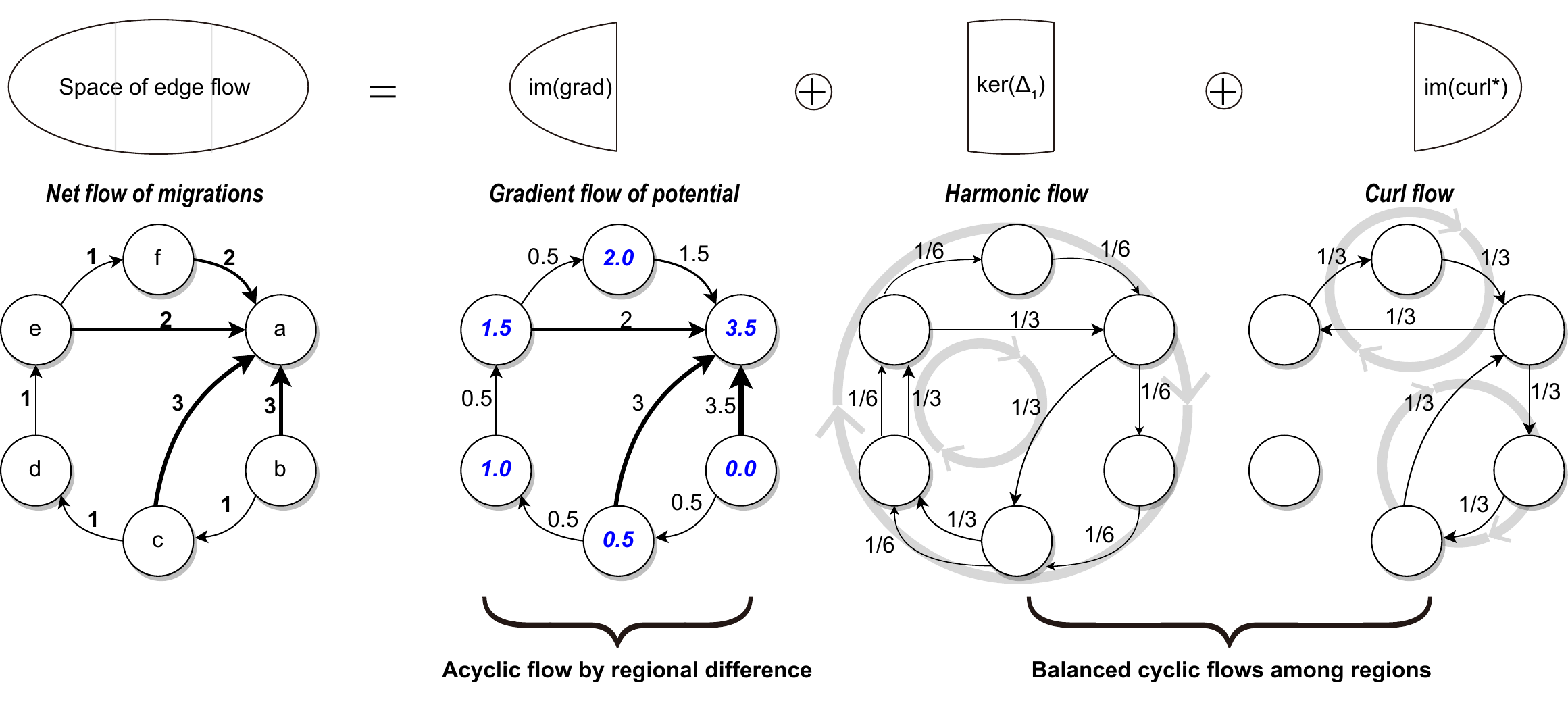}
  \caption{Hodge decomposition of a migration flow.
    The network in the left panel shows migration data among six locations.
    The black numbers of links denote the net movements between them.
    This flow is uniquely and orthogonally decomposed into gradient, harmonic, and curl flows in the combinatorial Hodge theory (top panel).
    Blue numbers in the gradient flow denote the potentials whose regional differences give the global acyclic part of flows and quantify the empirical liveability of each location.
    The other two flows are perpendicular to the acyclic flow: these flows are circular among locations, and incoming and outgoing fluxes are balanced.}
  \label{fig:illustration}
\end{figure}

Figure \ref{fig:illustration} serves as an explanatory illustration of the combinatorial Hodge decomposition.
There are six locations labelled $a,b,\cdots, f$ on the figure, and the net movements of migrations among them are depicted by a network (or graph) in the left panel.
This flow is uniquely decomposed into three distinct components: gradient, harmonic and curl.
In the figure, the black numbers of links denote the number of movements, and it is confirmed that the sum of the numbers of each link over three components equals that of the given flow (For example, at the link between $a$ and $b$, $3 = 3.5 - 1/6 - 1/3$).

The decomposed components exhibit distinct properties.
In particular, the gradient component is given by the differences in the scalar potential $s$, denoted by the blue numbers in the figure.
The higher the potential $s_i$, the more often location $i$ collects migrants from other locations.
Further, the gradient flow is acyclic and globally consistent.
This means that the accumulated gradient flows along different paths from one location to another are equivalent (e.g. the sum of the flows along the path $d \to c \to b \to a$  equals that of another path $d \to e \to f \to a$).
This is somewhat similar to the transitivity of preferences but stricter: values should be matched.
Therefore, the potential provides a consistent metric for regional liveability.
The other two components are circular.
Curl flow consists of triangular flows among triplets of locations,
and harmonic flow consists of other cyclic flows in more than three locations.
It is noted that these circular flows do not mean that individuals move to three or more locations. The flows represent aggregated movements at the population level.

To provide a general description, let us consider an undirected graph $G(V,E)$  with a set of vertices $V$ and a set of edges $E$. $N$ is the number of vertices.
We assign edge flow $Y$ to the edges.
$Y$ is an $N \times N$ matrix and its element $Y_{ij}$ is the flow from $i$ to $j$ and is skew-symmetric, $Y_{ij} = - Y_{ji}$.
Subsequently, the combinatorial gradient, curl and divergence are defined as follows \autocite{Jiang2011b}:
\begin{align*}
      (\text{grad}\, s)(i, j) &=  s_j-s_i \quad \text{for $\{i,j\} \in E$} , \\
    (\text{curl}\, Y)(i, j, k) &= Y_{ij} + Y_{jk} + Y_{ki}\quad \text{for  $\{i,j,k\}$}: \{i, j\}, \{j, k\}, \{k, i\} \in E ,\\
      (\text{div}Y)(i) &= \sum_{j \text{ s.t. } \{i,j\} \in E} Y_{ij},
\end{align*}
where $s$ ($\in \mathbb{R}^N$) denotes the potential to be introduced.
The space of the edge flow $\mathcal{Y}$ is orthogonally decomposed into images and kernels of the operators as follows (Top panel in Figure \ref{fig:illustration}):
\begin{align*} 
  \mathcal{Y}  = \text{im}(\text{grad})  \oplus \text{ker}(\Delta_1) \oplus  \text{im}(\text{curl}^*), %
\end{align*}
where ker($\Delta_1$) = ker(curl) $\cap$ ker(div) and $\text{curl}^*$ is the adjoint operator of the curl.
Because of the orthogonality of the decomposition, the harmonic and curl flows are divergence-free.
This means that incoming and outgoing migrants in these components are balanced. Therefore, the harmonic and curl components do not drive population change for each location.
Thus, this paper focuses on the potential of the gradient component, which is the driving force of unbalanced, acyclic migrations among regions.

Intuitively, gradient flow represents clear directions where one city tends to lose population consistently to another. In contrast, curl or harmonic flows represent balanced migrations, where no city loses or gains population overall.

The potential $s$ is defined as the solution to the following optimisation problem of weighted least squares:
\begin{align}
  \min_s \sum_{\{i,j\} \in E} W_{ij} \left[ \text{(grad $s$)}(i,j) - Y_{ij} \right]^2
  =
  \min_s \sum_{\{i,j\} \in E} W_{ij} \left[ (s_j - s_i) - Y_{ij} \right]^2 \label{eq:optimization}
\end{align}
where $W_{ij} (\in [0,1])$ is the weight of the pairwise comparison, which is primarily determined by the distance $d_{ij}$, as discussed later.
To find the potential $s$, this optimisation problem should be solved.
In general, minimizing errors of weighted least squares can be interpreted as a problem of finding the closest point of a subspace, which is solved by projecting onto the subspace \autocite{strang2023introduction}.
Similarly, the optimisation problem of Equation \eqref{eq:optimization} is the problem of finding the closest point to the given data $Y$ in the subspace of the edge flow and can be solved by an $l_2$-projection of $Y$ onto im(grad) \autocite{Jiang2011b}.
With a Euclidean inner product in space $\mathcal{Y}$, $\langle X,Y\rangle = \sum_{ \{i,j\} \in E} W_{ij} X_{ij}Y_{ij}$, the normal equation is given by
\begin{align}
  \Delta_0 s  = - \text{div} Y,  \label{eq:normal_equation}
\end{align}
where $\Delta_0$ is the graph Laplacian denoted by
\begin{align*}
  \left[ \Delta_0 \right]_{ij} =  \begin{cases}
    \sum_j W_{ij} \quad &\text{if $i = j$}\\
    -W_{ij}    \quad &\text{if $\{i,j\} \in E$}\\
    0    \quad &\text{otherwise}
  \end{cases},
\end{align*}
Finally, the potential $s$ is given by the minimal-norm solution of Equation \eqref{eq:normal_equation}
\begin{align}
  s  = - \Delta_0^{\dagger} \text{div} Y,  \label{eq:potential}
\end{align}
where $\dagger$ denotes the Moore-Penrose inverse.

\begin{figure}
  \includegraphics[width=\linewidth]{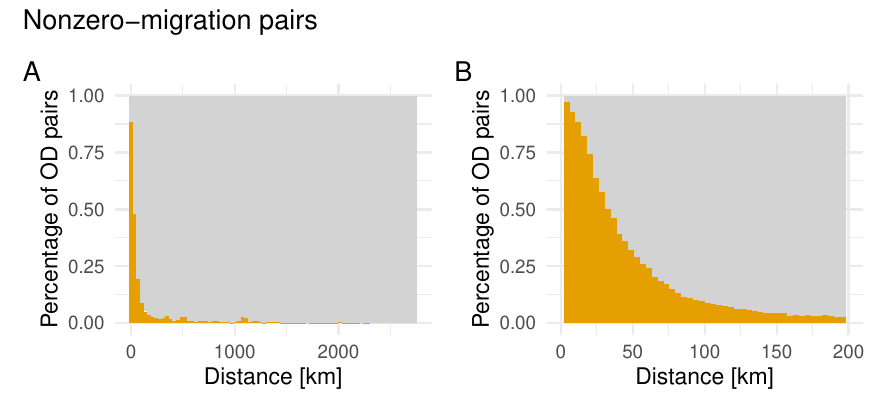}
  \caption{
Percentage of origin--destination (OD) pairs $(i,j)$ with nonzero migration ($M_{ij} > 0$ or $M_{ji} > 0$) as a function of road distance $d_{ij}$ in the Japanese migration OD matrix, $M$. 
    (\textbf{A}) The full range of road distance.
    (\textbf{B}) A focused range of $[0,200]$ km.
}
  \label{fig:zerotrip}
\end{figure}

How can this theory be applied to migration flow datasets?
To determine this, for a given OD matrix $M$ of the migration flow, we specify the edge flow $Y$ by its net flow, as follows:
\begin{align*}
  Y = M - M^{\top}. 
\end{align*}
Its element $Y_{ij}$ denotes the net flow from $i$ to $j$ and satisfies the necessary skew-symmetric condition in the theory. 

The distances between locations matter for weights $W$ and edges $E$.
In the definition of the potential using the optimisation problem \eqref{eq:optimization},
the regional difference in potential, $s_j - s_i$, is compared with the net flow  $Y_{ij}$ only for  pairs in the set of edges $E$ and the comparison is weighted by $W_{ij}$.
A naive method to determine $E$ and $W$ chooses pairs of locations $(i,j)$ as $E$ if there is a migrant between them, that is, $M_{ij} >0$ or $M_{ji} > 0$, with equal weights, $W_{ij} =1$.
Alternatively, here, we consider distance-based weighting in a data-driven manner.
Let us examine the dependence of migration flow on distance.

Figure \ref{fig:zerotrip} shows the percentage of nonzero migration pairs ($M_{ij} > 0$ or $M_{ji} > 0$)  as a function of road distance $d_{ij}$ in the Japanese migration dataset (see `Methods' Section for details). %
We found that the percentage quickly decreased as distance increased. Even at the relatively short distance $d_{ij} \sim $ 40 km, approximately half of the pairs had no migrants.

What does zero migration mean?
Migration incurs considerable costs.
Zero migration indicates that the cost is higher than the returns to migration owing to regional differences.
When the cost is enormous owing to long-distance separation, the zero migration is common even if the regional difference is nonnegligible, that is, $|s_j -s_i| \gg 0$.
In this case, in Equation \eqref{eq:optimization} the squared residual $ \left[\text{(grad $s$)}(i,j) - Y_{ij} \right]^2$ = $\left[ (s_j - s_i)  - Y_{ij} \right]^2$ = $\left[ s_j - s_i \right]^2$ should be penalised with  almost zero weight ($W_{ij} \sim 0$).
On the other hand, if the cost is relatively small over a short distance, zero migration is not common and implies that the regional difference $s_j -s_i$ is sufficiently small.
In this case, the residual should be largely weighted ($W_{ij} \sim 1$), penalising the difference of the potential $(s_j - s_i)^2$. 
Based on these considerations, we evaluate the weight $W_{ij}$ of the zero-migration pair $(i,j)$, by measuring how commonly the pairs of locations could have some migrants overcoming the cost barrier by distance $d_{ij}$, $C(d_{ij})$.
Instead of predetermined distance deterrence functions, such as exponential or power-law forms, $C(d)$ is empirically determined as a function of distance $d$, using the histogram in Figure \ref{fig:zerotrip}.
In contrast to the zero-migration situation, when there are some trips between a pair of locations, the weight is simply set to be 1 to minimise the residual, $\left[ (s_j - s_i)  - Y_{ij} \right]^2$.
In summary, the weight $W$ is given by:
\begin{align*}
  W_{ij} = \begin{cases}
    1 \quad  &\text{non-zero migration pair ($M_{ij} >0$ or $M_{ji}  > 0$)} \\
    C(d_{ij}) \quad & \text{zero migration pair ($M_{ij} = M_{ji} = 0$)}
\end{cases}.
\end{align*}
For the set of edges $E$, we choose the pairs of positive weights, $\{ (i,j) \mid W_{ij} > 0\}$.

\subsection{The potential as an extension of net migration and regional utility }
\label{sec:methods_completegraph}
Net migration is a traditional measure of migration flow and is given by the difference between the incoming and outgoing fluxes at each location $i$, $\sum_{j \neq i} M_{ij} - \sum_{j \neq i}M_{ji}$.

How is the proposed potential related to net migration?
To determine this, under a specific condition, wherein $E$ is a complete graph with equal weights $W_{ij} =1$,
Equation \eqref{eq:potential} is simplified as 
\begin{align}
  s = - \frac{1}{N} \text{div} Y = \frac{1}{N} \left[ \sum_{j \neq i} M_{ji} - \sum_{j \neq i} M_{ij} \right], \label{eq:potential_completegraph}
\end{align}
where $N$ denotes the number of locations. 
This is equivalent to the net migration divided by $N$.

The derived equation has two messages:
First, net migration is supported by mathematical reasoning based on the combinatorial Hodge theory.
Second, the potential introduced in this study can be interpreted as an extension of net migration.
The specific condition ($E$ is a complete graph with equal weights $W_{ij} =1$), from which net migration is derived, ignores the distance factor of migration; that is, all locations are connected with equal weights.
Therefore, instead, we consider spatial frictions in migration flows between locations using $W$ and $E$.
The potential introduced in this study can thus be interpreted as an extension of net migration, explicitly considering the distance factor in the weights $W$.

Regional utility differential estimated from migration datasets is another related metric to the potential. As shown in the detailed derivations presented in the `Methods' Section, %
the regional utility can be interpreted as a potential in the combinatorial Hodge theory, under a specific migration flow indicator without distance factors.

\subsection{Empirical liveability for young migrants in Japan}
\label{sec:potential} 
In the following two sections, we demonstrate how the proposed method can be used in the context of practical analysis.
As a case study, we focused on two types of migrants between Japanese municipalities: WoRA and FwSC. %
In this section, we first explain the migration dataset to be analysed. Second, we present the potential, representing  empirical liveability, for the two types of populations. Third, we characterise the municipalities for these groups, using the combinatorial classification by statistical test to check the significant sinks and sources of those migration flows.

We used datasets from the `Report on Internal Migration in Japan Derived from the Basic Resident Registration'. 
The datasets summarise the number of migrants between municipalities from the beginning to the end of each year.
In this study, we examined the report in 2019 to exclude the impact of the COVID-19 pandemic.
There are 8.3 million migrants in these municipalities.
We selected two specific populations from the datasets: FwSC and WoRA.
The datasets are categorised according to age (every ten years) and sex.
However, information about family structure is not provided.
Therefore, we could not select adults who have small children directly.
Alternatively, as proxies for FwSC, we selected children aged 0--9 because they are expected to migrate with their parents.
We also considered a population of WoRA by selecting women aged 20--39.
According to the Center for Disease Control and Prevention in the US, WoRA are defined as women aged 18--44 years.
Owing to the limitations of the datasets, we chose the largest subset (20--39 years) as per the definition.

\begin{figure}
    \centering
  \includegraphics[width=\linewidth]{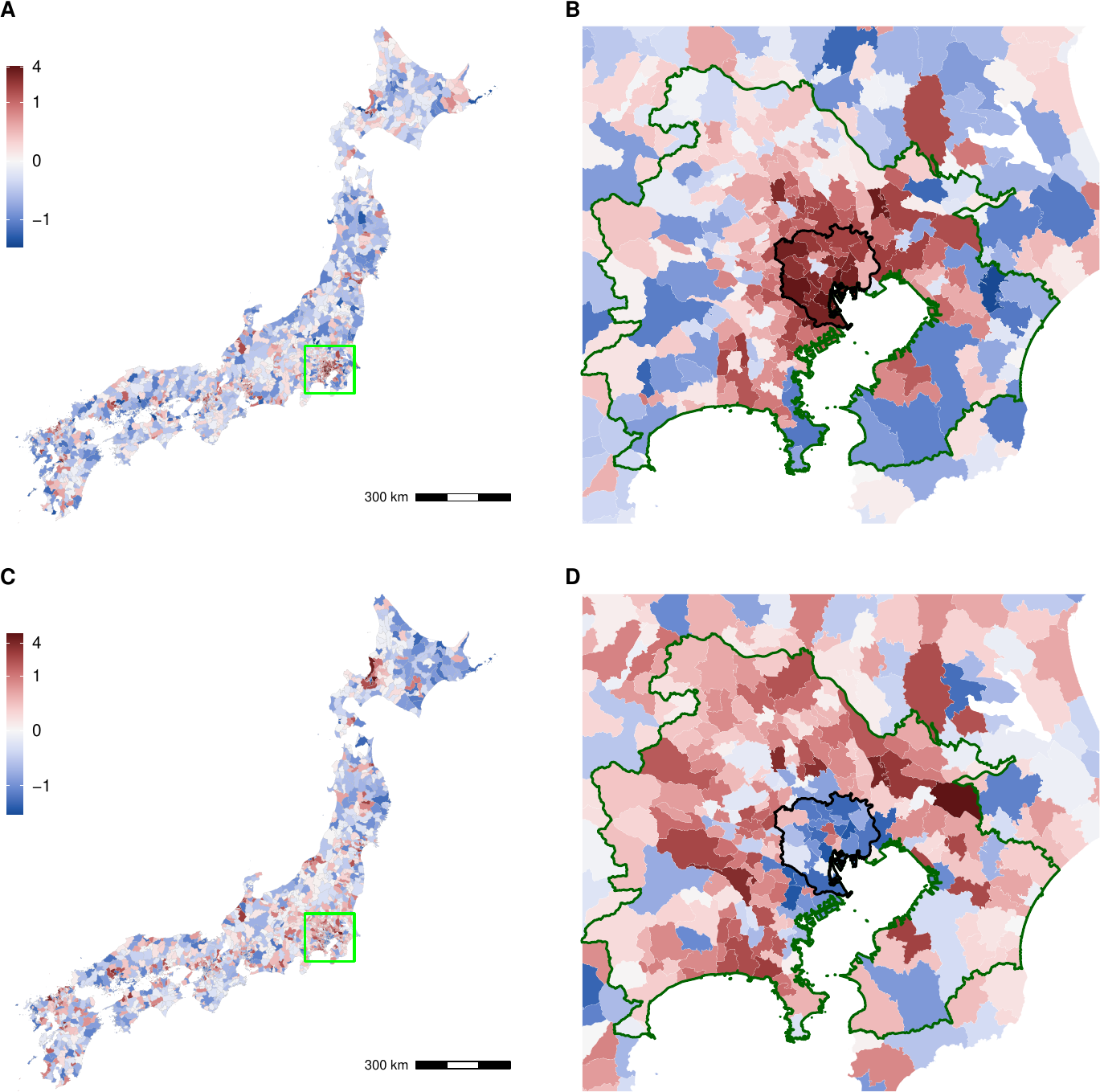}
  \caption{
    (\textbf{A,B})
    The potential landscape of migration flows for \textit{women of reproductive age} in the entire target area in Japan (A) and the focused area around Tokyo (B), shown in a green rectangle in the left panel.
    The unit of the potential analysis is municipality in Japan.
    The black boundary illustrates the area of the special wards of Tokyo, known as the central area of the metropolis. 
    The green boundary indicates Tokyo Metropolitan Employment Area as an alternative to an official metropolitan area.
    (\textbf{C,D})
    Same as A, B, but for \textit{families with small children}.
    Larger images of Panels A and C are in Supplementary Information.
  } 
  \label{fig:potential}
\end{figure}

Figure \ref{fig:potential} provides an overview of the potentials.
The mean of the potentials obtained by Equation \eqref{eq:potential} is zero, $\sum_i s_i = 0$.
Therefore, on average, municipalities with a positive potential attract migrants from other municipalities, whereas those with a negative potential do not, on average, attract migrants.

The landscape of the potential differs crucially between the two subgroups.
For WoRA, higher-potential areas were more concentrated in the central areas, such as the special wards of Tokyo (shown as a black boundary in Fig. \ref{fig:potential}B): the special wards of Tokyo consist of 23 wards. The whole area of these wards was former Tokyo city until 1943, and the area is known as the central area of Tokyo metropolis.
For FwSC, the potentials around the Tokyo's central areas were negative (Fig. \ref{fig:potential}D), and higher-potential areas were often seen outside the central areas, within Tokyo Metropolitan Employment Area (illustrated as a green boundary in Fig. \ref{fig:potential}C).
It is noted that the metropolitan area of Tokyo is not officially defined by the Statistics Bureau of Japan. Alternatively, researchers defined the Metropolitan Employment Areas as functional urban areas consisting of urban cores and the surrounding municipalities that exhibit strong commuting patterns from the latter to the former \autocite{Kanemoto2002UEA}. We used the 2015 standards obtained from \url{https://www.csis.u-tokyo.ac.jp/UEA/}.

\begin{figure}
    \centering
    \includegraphics[width = 0.5 \linewidth]{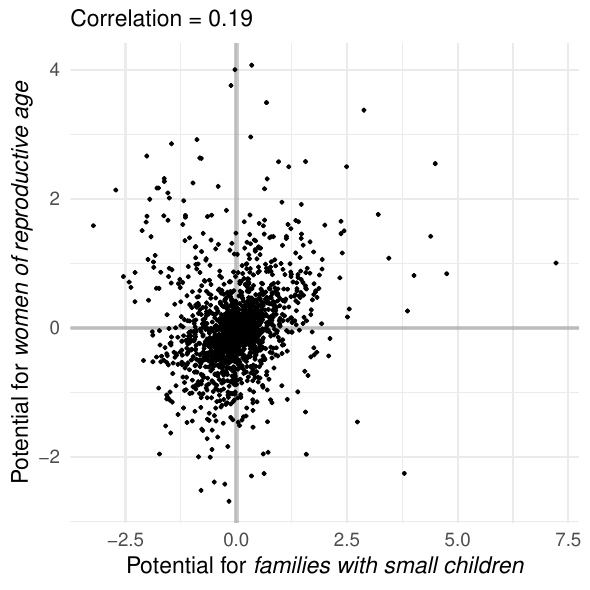}
    \caption{
Comparison between the \textit{women of reproductive age} and \textit{families with small children} groups by their potentials. A point corresponds to a municipality.
}
\label{fig:scattering}
\end{figure}

We also confirm that the potentials of these groups were very weakly positively correlated or almost uncorrelated (Figure \ref{fig:scattering}), indicating that these groups will have different preferences for liveable cities.
It is noted that the correlations are different between urban and rural areas.
Municipalities in Japan are legally classified into cities (including wards), towns, and villages. We separated the municipalities into two groups: cities and others (towns and villages).
We found that the former group (cities) still showed a very weak positive correlation (=0.14).
On the other hand, the latter group (towns and villages) showed a moderate positive correlation (=0.43).

\begin{figure}
    \centering
    \includegraphics[width = \linewidth]{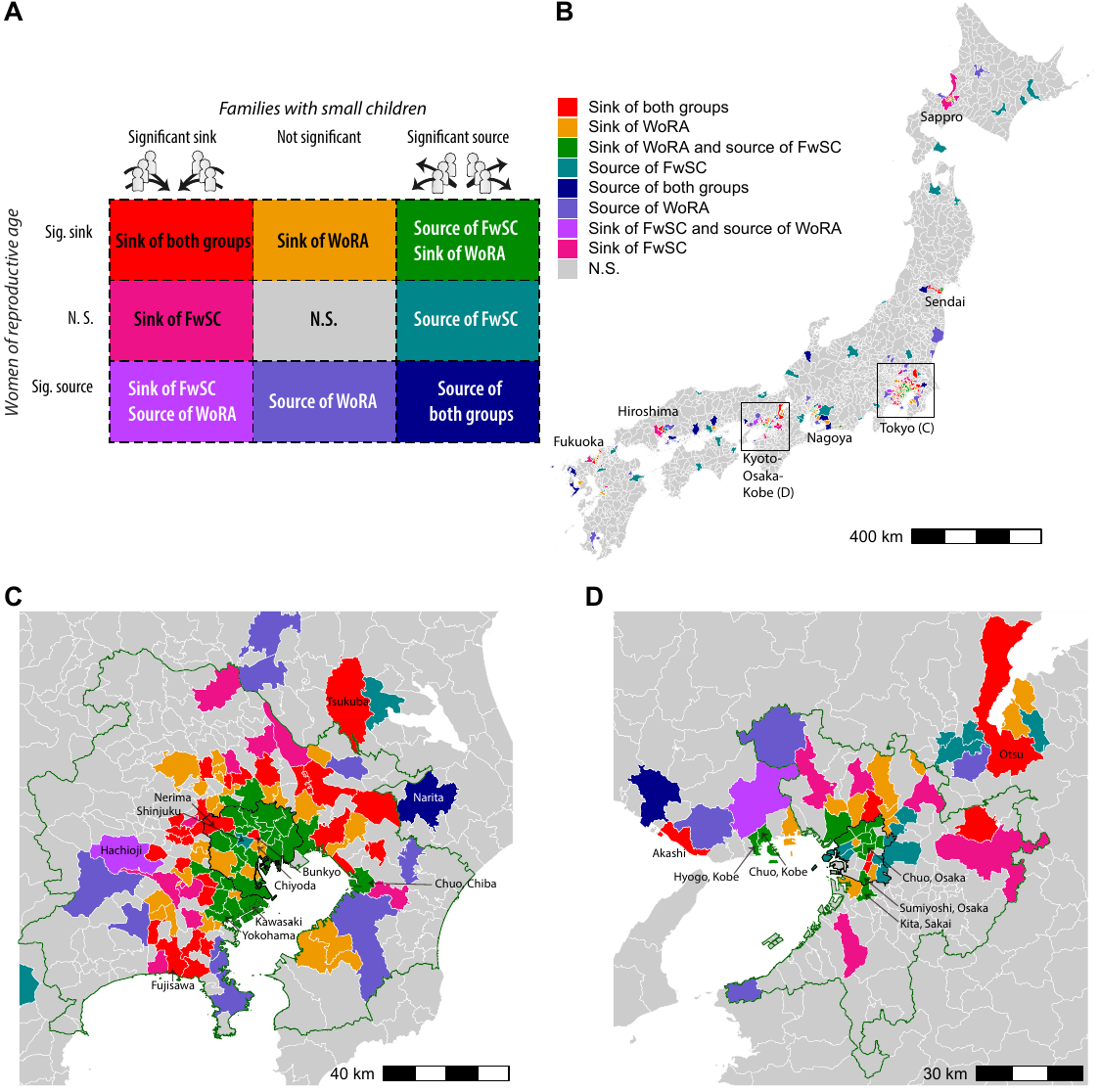}
    \caption{
      (\textbf{A})
Combinatorial classification by statistical testing for the migrations of two specific populations: \textit{women of reproductive age} (WoRA) and \textit{families with small children} (FwSC).
      (\textbf{B})
The map of the classified municipalities in the entire target area in Japan.
Significant municipalities of migrations are indicated by colours, according to the classification scheme shown in A.
      (\textbf{C})
      A focused map of a Tokyo area, indicated by a rectangle in panel (B).
The black boundary illustrates the area of the special wards of Tokyo, known as the central area of the metropolis.
The green boundary indicates Tokyo Metropolitan Employment Area as an alternative to an official metropolitan area.
      (\textbf{D})
      A focused map of a Kyoto-Osaka-Kobe (\textit{Keihanshin}) area, indicated by a rectangle in panel (B).
The black boundary illustrates the area of Osaka city as the central area, and the green boundary indicates Osaka Metropolitan Employment.
}
\label{fig:classification}
\end{figure}

In Figure \ref{fig:scattering}, municipalities in different quadrants have distinct properties.
Municipalities in the first quadrant were attractive to both groups.
Those in the second quadrant were attractive to WoRA, but not attractive to FwSC.
Those in the third quadrant were unattractive to both groups.
Those in the fourth quadrant were attractive to  FwSC, but not attractive to WoRA. 

To identify these distinct properties of the municipalities,
we introduced the statistical detection of significant sinks and sources of migration flows, which are tested to have a higher or lower potential than a counterfactual null model \autocite{aoki2023identifying}.
In the null model, flow $Y$ is randomly shuffled among $E$, while the weight matrix $W$, determined by the distances between municipalities, is not randomized.
We performed a permutation test with a Monte Carlo simulation and a multiple comparison test under the control of the false discovery rate  $\alpha$ (= 0.05).
See  Aoki \textit{et al.} \autocite{aoki2023identifying} for further details.

Figure \ref{fig:classification}(A) illustrates the classification scheme for significant migration flow locations.
Municipalities are classified as significant sinks, significant sources and others (not significant) for the migration of FwSC.
The municipalities are also classified into three cases of WoRA.
Combining these classifications, there are three $\times$ three types of municipalities.
For simplicity, we omit the word `significant' in the significant sink and significant source classification labels, and use \textit{sink} and \textit{source} only.

Figure \ref{fig:classification}(B) shows a map of the classified municipalities in the target area.
The `sink of both groups' (red-coloured) is the attractive municipality for the two specific populations of WoRA and FwSC.
They were located in the major metropolitan areas of Japan, such as \textit{Tokyo}, \textit{Kyoto-Osaka-Kobe}, \textit{Fukuoka}, \textit{Hiroshima}, and \textit{Sendai}.
Significant sinks of either group (coloured in orange or pink) were observed in these and other metropolitan areas such as \textit{Nagoya} and \textit{Sapporo}.
On the source side (the municipalities where people leave), the detected municipalities (coloured in blue tones) were widely seen in several municipalities in Japan, including major metropolitan areas.

Figure \ref{fig:classification}(C) shows the rectangular area around the Tokyo metropolitan area.
The green boundary indicates Tokyo Metropolitan Employment Area, mentioned above, as the entire metropolis area.
Inside the green boundary, the black border shows the special wards of Tokyo, which is the core of the metropolis.
In this central area, most municipalities were `sink of WoRA and source of FwSC' (green-coloured).
These green-colored municipalities appeal differently to these two groups, evidenced by WoRA migrating into them and FwSC migrating out.
In this central area, a few municipalities were `sink of WoRA' (orange), including \textit{Chiyoda}, which is known as the economic and political centre of Japan.
\textit{Nerima} and \textit{Bunkyo} were `sink of both groups' (red). 
\textit{Shinjuku} was `source of FwSC' (blue-green).
Other green-type municipalities were observed in places contiguous to the centre, such as several wards in the cities of \textit{Kawasaki} and \textit{Yokohama}, known as the subcentres of the Tokyo metropolitan area.
Another green type was seen at \textit{Chuo} ward in \textit{Chiba}, one of Tokyo's subcentres.
In the surroundings of the green-dominant area, there are many municipalities of `sink of both groups' (red) and the sink of either of the groups (pink or orange).
Therefore, migrants detected these locations as liveable cities.
Notably, municipalities in the surrounding areas were not always detected as sinks.
Moreover, in areas that are over 90km away from the centre, some municipalities such as \textit{Tsukuba} and \textit{Fujisawa} were detected as `sink of both groups' (red), located around the boundary of Tokyo Metropolitan Employment Area.
On the other hand, several municipalities around the boundary were detected as the `source of WoRA' (bluish-purple).
In addition, `sink of FwSC and source of WoRA' (purple) was detected at \textit{Hachioji}.
This was another type of municipality evaluated with the opposite preference for these groups.
`Source of both groups' (dark blue) was observed at \textit{Narita}. 

Figure \ref{fig:classification}(D) shows a rectangle area in Kyoto-Osaka-Kobe (\textit{Keihanshin}), which is Japan's second-largest metropolitan area.
The green boundary indicates Osaka Metropolitan Employment Area and the black border shows the city of \textit{Osaka} as its central area.
In the north area of the centre, green-type municipalities were dominant, including \textit{Chuo} (central) ward.
In the south area of the centre, \textit{Sumiyoshi} ward in \textit{Osaka} and \textit{Kita} ward in \textit{Sakai} city were detected as green-type municipalities.
In the surroundings of these green-type areas, several municipalities were the sinks of the two groups (red, orange, or pink).
Other green-type municipalities were observed in the \textit{Chuo} and \textit{Hyogo} wards in \textit{Kobe} city as one of the main sub-centres in the \textit{Keihanshin} metropolis.
In areas that are over 90km apart from the centre, some municipalities, such as \textit{Otsu} and \textit{Akashi}, were detected as `sink of both groups' (red).

These observations imply the distinct preferences of the two groups, leading to their spatial segregation.
Roughly speaking, WoRA preferred the central area and its surroundings in metropolises to the suburbs that are far from the centre.
Conversely, FwSC preferred several specific municipalities in the surrounding areas to the central areas.
In the next section, we attempt to answer the following questions:
(1) What factors determine the preferences of these people?
and (2) what are the differences between attractive and unattractive municipalities? 
For this, we performed a regression analysis to examine possible factors.

\subsection{What factors are important for liveability?}
\label{sec:regression} 
\begin{table}[!htbp] \centering 
  \caption{Linear regression analysis using the potentials as the target variable.} 
  \label{tbl:regression} 
\normalsize 
\begin{tabular}{@{\extracolsep{5pt}}lD{.}{.}{-3} D{.}{.}{-3} } 
\\[-1.8ex]\hline 
\hline \\[-1.8ex] 
 & \multicolumn{2}{c}{\textit{Dependent variable:}} \\ 
\cline{2-3} 
\\[-1.8ex] & \multicolumn{1}{c}{Women of reproductive age} & \multicolumn{1}{c}{Families with small children} \\ 
\hline \\[-1.8ex] 
 Population & -0.035 & -2.083^{***} \\ 
  HabitableArea & -0.336^{***} & -0.029 \\ 
  Income & -0.052 & -0.220^{*} \\ 
  LandPrice & -0.006 & 0.011^{*} \\ 
  LFPR & 0.080^{***} & -0.002 \\ 
  UnemploymentRate & 0.038 & 0.051 \\ 
  Manufacturing & -0.006 & 0.0001 \\ 
  BusinessService & -0.004 & -0.012 \\ 
  HumanCapital & 0.056^{***} & 0.042^{***} \\ 
  HousesFarFromPreschool & 0.005^{*} & -0.002 \\ 
  STRatio\_Elementary & 0.061^{***} & 0.083^{***} \\ 
  Library & 0.011 & 0.037^{*} \\ 
  LargeScaleRetailers\_10km & 0.343^{***} & -0.009 \\ 
  MeanCommutingTime & -2.324^{***} & 1.847^{***} \\ 
  HousingStarts & 0.221 & 0.571^{**} \\ 
  VacancyRate & 0.004 & -0.001 \\ 
  OwnerOccupied & 0.011^{*} & 0.058^{***} \\ 
  SewageTank & -0.255 & 0.693 \\ 
  HospitalPerCaptia & -1.319 & -1.410 \\ 
  TheftPerCaptia & 0.012 & 0.034 \\ 
  CityClass & 0.021 & -0.044 \\ 
  Constant & -6.160^{***} & -6.590^{***} \\ 
 \hline \\[-1.8ex] 
Observations & \multicolumn{1}{c}{1,212} & \multicolumn{1}{c}{1,212} \\ 
R$^{2}$ & \multicolumn{1}{c}{0.450} & \multicolumn{1}{c}{0.357} \\ 
Adjusted R$^{2}$ & \multicolumn{1}{c}{0.436} & \multicolumn{1}{c}{0.341} \\ 
\hline 
\hline \\[-1.8ex] 
\textit{Note:}  & \multicolumn{2}{r}{$^{*}$p$<$0.05; $^{**}$p$<$0.01; $^{***}$p$<$0.001} \\ 
                & \multicolumn{2}{r}{Each model includes region dummies (region fixed effects).}\\
                & \multicolumn{2}{r}{The coefficients of region dummies are included and shown in Figure \ref{fig:regression_region}.} \\ 
\end{tabular} 
\end{table}

Using the potentials for WoRA and FwSC as dependent variables, we performed a regression analysis to examine the factors that determine the liveability score (Table \ref{tbl:regression}).
To compare these two cases, the potential was standardised using the standard deviation of all municipalities. 
We chose several explanatory variables related to income, rent, price and several types of amenities, following previous studies \autocite{blomquist1988new, diamond2016determinants, glaeser1992growth, glaeser1995economic, glaeser2001consumer, glaeser2004rise, glaeser2014cities, shimizu2014urban} (For details, please see `Datasets of regression analysis' in the Supplementary Information).
The descriptive statistics are summarised in Supplementary Table 1.
Some variables are available only for municipalities with a population large enough to be included in specific Japanese Census surveys.
These liveability-related variables are inevitably correlated to each other.
We checked that the correlations were less than 0.8 and the generalised variance-inflation factors (GVIFs) below a standard rule of thumb: $GVIF^{1/df} < 10$, where $df$ is the degree of freedom to adjust for the dimension of the confidence ellipsoid \autocite{fox2018companion}.
We also checked the robustness of the obtained result by removing the non-significant variables in Table \ref{tbl:regression} (see Supplementary Table 2) and confirmed that the significant variables in Table \ref{tbl:regression} were not altered to be non-significant, except for two cases whose criteria were $p < 0.05$: \textit{Income} and \textit{Land Price} for FwSC. Therefore, we focused on significant variables with the criterion $p < 0.01$.

First, \textit{Population} (unit is 1 million) and \textit{Habitable Area} (unit is 100 km$^2$) are the controls of the regression analysis.
The dependent variable --- potential --- induces a gradient flow of migrants by difference and is relevant to the size of the municipality.
Controlling for the municipality size by population and habitable area allows us to examine the regression coefficients of the other variables.
Yet, the estimated coefficients for these control variables are themselves informative.
The coefficient for population was significantly negative ($p < 0.001$) for FwSC,
whereas the habitable area of the municipality had a negative effect ($p < 0.001$) on the potential for WoRA. 
These observations might suggest that WoRA prefer higher-density populations, while FwSC prefer lower-density areas.

\textit{Income} (unit is 1 million yen) represents the taxable income, calculated using the aggregated income of a municipality divided by the number of taxpayers.
\textit{Land Price} (unit is $10^4$ yen per square meter) is a price variable, giving the price of land amount per square meter for residential areas.
In this analysis, we found significant effects of these variables for FwSC,
but they were unstable for the change of explanatory variables (Supplementary Table 2).

\textit{Labour Force Participation Rate (LFPR)} and \textit{Unemployment Rate} represent labour market conditions, given by percentages. 
LFPR had a significant effect ($p < 0.001$) for WoRA. 
The unemployment rate had no significant effect.

The variables \textit{Manufacturing} and \textit{Business Service} capture industry composition \autocite{glaeser1995economic,glaeser2004rise}, with each defined as the share of employees employed in the corresponding sector.
We found no significant effects on these variables.

\textit{Human capital} has been argued to be a relevant indicator of urban growth \autocite{glaeser2004rise, glaeser2014cities}.
We evaluated it as the percentage of graduates over the age of 25 years.
We found that human capital had a significantly positive effect ($p < 0.001$) for both groups in our study.
This variable would be a proxy for skilled and adaptive workers relevant to the productivity and economic growth of cities \autocite{glaeser2004rise}.
From a consumption viewpoint, the variable could be a proxy for community building and the quality of public institutions, such as public schools and museums \autocite{diamond2016determinants}.
Although our result is consistent with the causal interpretation of previous studies mentioned above, human capital could also be influenced by confounding factors not included in our explanatory variables. The presence of administrative facilities, transport infrastructure and business districts, for example, can attract corporate headquarters, which in turn draw skilled workers. 

We also included educational amenities \autocite{diamond2016determinants} in the analysis, specifically \textit{Houses far from preschool}, \textit{ST Ratio Elementary} and \textit{Library}.
\textit{Houses far from preschool} is the percentage of houses whose nearest preschool is over 1km apart, and it had a positive effect ($p < 0.05$) for WoRA.
\textit{ST Ratio Elementary} is the student--teacher ratio in elementary schools.
For both groups, this variable had a significant positive effect ($p < 0.001$).
This means that a higher student--teacher ratio is evaluated positively.
However, a low value was considered beneficial.
In the Japanese educational system, an elementary class has a maximum of 35 students.
For large schools, almost all classes have the maximum number of students.
However, small schools in depopulated areas often have classes with few students. %
Therefore, this ratio can serve as a proxy for school size.
\textit{Library} (unit is one) is the number of libraries in each municipality, and had a positive effect ($p< 0.05$) only for FwSC.

Drawing on Glaeser \textit{et al.}'s (2001) consumer-city view \autocite{glaeser2001consumer}, which highlights the importance of consumption amenities, this study also examines the effect of large-scale retailers on the liveability of cities.
In the Japanese census, retailers with more than 50 employees are categorised as large-scale retailers.
\textit{Large Scale Retailers 10km} (unit is 100) is the number of such retailers in each municipality and the neighbouring municipalities.
We discounted the number of retailers in  neighbouring municipalities at distance $d$,
using the distance-deterrence function $f(d) = \exp(d/d_0)$ with $d_0$ = 10km.
The variable showed a significant positive effect ($p < 0.001$) only for WoRA, suggesting that consumption amenities are vital to them.

\textit{Mean commuting time} (unit is one hour) evaluates the accessibility of municipalities using commuting time averaged by households mainly supported by an employee.
We found that mean commuting time had a significantly negative effect ($p < 0.001$) for WoRA but a positive effect ($p < 0.001$) for FwSC.
Further, mean commuting time was closely related to the distance from metropolises' centres and subcentres, and that the negative (positive) coefficient for WoRA (FwSC) was consistent with the observed preference for the centres (suburbs).

Several statistics concerning housing, which is an essential factor in migration, were included.
These include \textit{Housing starts}, \textit{Vacancy rate}, \textit{Owner occupied} and \textit{Sewage tank}.
\textit{Housing starts} (unit is $10^3$ houses), which refers to the number of houses under construction, had a significant positive effect ($p < 0.01$) on FwSC.
It is noted that the construction of more houses could be an outcome of an inflow of FwSC.
\textit{Vacancy rate}, the percentage of unoccupied houses over total houses, had no significant effect.
\textit{Owner occupied}, the percentage of owner-occupied houses over the total, had a significant positive effect on WoRA ($p < 0.05$) and FwSC ($p < 0.001$).
\textit{Sewage tank}, the percentage of residents with proper sewage disposal facilities, showed no significant effect.
These observations suggest that FwSC prefer residential areas with newly built houses and those occupied by their owners.
By contrast, WoRA could not have clear, significant preferences. %

\textit{Hospital per capita} (unit is one per $10^3$ residents), which refers to the number of hospitals, describes the healthcare environment in each municipality.
It had no significant effect, contrary to the expectation that medical institutions will attract migrants.

Since crime is considered a relevant factor influencing the liveability of cities \autocite{diamond2016determinants}, it was also analysed.
Specifically, \textit{Thefts per capita} (unit is one event per $10^3$ residents), which refers to the number of thefts per capita, based on publicly available data on Japanese crime (\url{https://www.npa.go.jp/toukei/seianki/hanzaiopendatalink.html}), showed no significant effects for either group.
The \textit{City class} variable is a dummy indicating whether a city is designated as either a `core city' or a `designated city' under Japanese law. These city types are delegated a range of administrative functions from the prefectural government. We included this variable to capture the effect of a city's administrative status; however, its effect was not statistically significant.

\begin{figure}[tb]
    \centering
    \includegraphics[width = \linewidth]{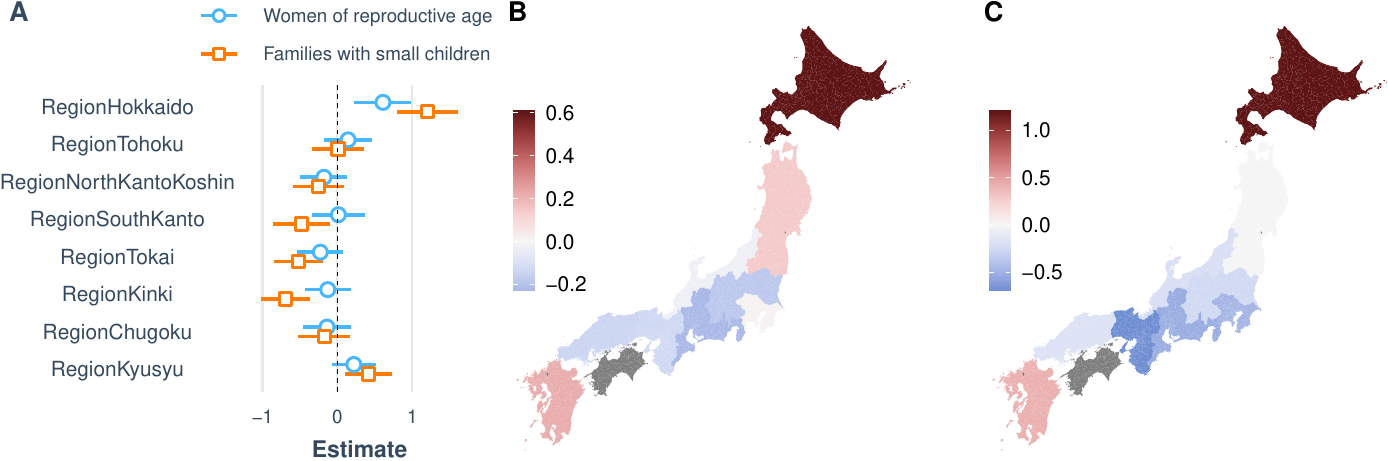}
    \caption{
      (\textbf{A})
    Estimated coefficients of region dummies in the linear regression analysis shown in Table \ref{tbl:regression}.  \textit{Shikoku} region was the reference level.
      (\textbf{B})
      The map of the estimates for \textit{women of reproductive age}.
      (\textbf{C})
      The map of the estimates for \textit{families with small children}.
}
\label{fig:regression_region}
\end{figure}

We also included a categorical variable of Japanese regions in the regression. These regions are based on the definitions provided by the Statistics Bureau of Japan \autocite{region}.
To convert the categorical variable into dummies, we set the baseline to the \textit{Shikoku} region, which has no core or government ordinance cities.
Figure \ref{fig:regression_region} shows the estimated coefficients depicting regional differences in Japan.
Among these regions, the \textit{Hokkaido} region had more positive coefficients than the other regions for FwSC.

In summary, under the criterion of $p < 0.01$, both groups had significant coefficients with the same sign for \textit{Human capital}(+) and \textit{ST ratio elementary}(+), but with opposite signs for \textit{Mean commuting time}.
Several variables were significant in only one of the groups. 
WoRA had significant coefficients for \textit{Habitable area}(-), \textit{LFPR}(+), and \textit{Large scale retailers 10km}(+),
while FwSC had significant coefficients for \textit{Population}(-), \textit{Housing starts}(+), and \textit{Owner-occupied}(+).
These groups have different, and sometimes conflicting, preferences for liveable cities.

\section{Discussion}
\label{sec:discussion} 
In this study, we proposed empirical liveability based on people's votes with their feet as a concrete KPI for local governments.
Although such liveability is hidden in the migration data that describes the pairwise comparisons of locations by migrants, we revealed that the liveability can be extracted using the `potential' in the combinatorial Hodge theory, directly calculated from migration data without any additional sources.
This mathematical framework allows us to reformulate the net migration and regional utility reported in previous studies, revealing that these measures can be interpreted as potentials with an interval scale, giving globally consistent rankings, while these measures ignore distance factors in migration.

We demonstrated this method in the practical case of Japanese migration flows between municipalities and evaluated the liveability scores for two demographic groups: FwSC and WoRA. %
We found that the empirical liveability of municipalities for these groups was almost uncorrelated.
Then, we categorised the municipalities by identifying the significant sources and sinks of both migration groups.
In addition to the observed overall migration pattern that revealed WoRA preferred central areas and FwSC preferred surrounding areas,
the classification of individual municipalities clarified their characteristics.
For example, the Bunkyo and Nerima wards in Tokyo attracted both populations despite being located in dense city centres. On the other hand, not all municipalities in the suburbs attracted FwSC.
These observations led us to further studies on the underlying factor of the evaluated potential, to facilitate evidence-based policymaking in each municipality. Therefore, we performed regression analyses using the scores as dependent variables to examine the factors determining the observed differences in their preferences.
Several factors were differently related to the scores of the migration groups.
In particular, commuting time had significantly opposite coefficients, and several variables were significant in only one of the groups.

These obtained results align with the previous studies in the regional science and urban economics literature.
Glaeser \& Kahn \autocite{GlaeserKahn2004} argue that increased demand for land --- driven by factors such as larger family size --- leads to urban sprawl, as households move outward in search of more affordable land.
In Abe's (2011) model \autocite{abe2011family}, a higher valuation of the wife’s leisure time, including time spent on parenting, reduces the likelihood of her full-time employment and lowers overall household income, which in turn strengthens the preference for suburban living where land is relatively inexpensive.
Using Japanese Census data for the Tokyo metropolitan area, Sakanishi\autocite{sakanishi2024empirical} finds that commuting time tends to increase with family size or the number of children, implying a preference among families with children for suburban areas within the Tokyo metropolitan area.
Further, robust positive coefficients of human capital are consistent with the empirical studies on urban growth \autocite{glaeser2004rise,glaeser2014cities}.

In summary of the case study, the results imply that the two groups, FwSC and WoRA, have different and sometimes conflicting preferences concerning liveability.
This suggests that there is no one-size-fits-all `liveable municipality.'
While this finding may seem intuitive, it offers crucial insights, as local governments in Japan --- in response to accelerating depopulation --- often adopt uniform policies \autocite{Takeda2018} to encourage migration to their regions.
Therefore, municipalities should identify specific demographic targets and tailor transportation, housing, and public service packages accordingly.

While the proposed method is explored here only in the context of Japanese migration,
it is applicable to other situations as well.
Nevertheless, the following limitations and considerations are important to note before applying this method.

First, in the proposed potential analysis, cyclic movements in harmonic and curl components were not the main focus, because in such movements, the numbers of incoming and outgoing migrants are balanced, and the movements do not influence population changes.
However, cyclic movements can be important for understanding migration patterns in urban studies, by inducing some demographic changes, such as through gentrification.
Selecting related subpopulations, for example selecting by economic class, could potentially address the demographic changes in the proposed analysis.

A second point concerns the selection of the datasets for specific populations.
The proposed method allows us to characterise places for specific groups according to attributes, such as ethnicity, education, skill, income, and others.
The spatial classification by a breakdown into specific demographic groups can be utilised to identify potential targets for place-based policies.
The selection is mathematically supported in the proposed method, because
potential $s$ is a linear map of a given net flow $Y$.
Therefore, the potential for all migrants equals the sum of those for subgroups:
For example, when net flow $Y$ is divided into three ethnicity subgroups, $Y = Y^{\mathrm{A}} + Y^{\mathrm{B}} + Y^{\mathrm{C}}$, the sum of potential for each group equals the total potential: 
$
s  = - \Delta_0^{\dagger} \text{div} Y = - \Delta_0^{\dagger} \text{div} (Y^{\mathrm{A}} + Y^{\mathrm{B}} + Y^{\mathrm{C}})
= - \Delta_0^{\dagger} \text{div} Y^{\mathrm{A}}  
- \Delta_0^{\dagger} \text{div} Y^{\mathrm{B}}  
- \Delta_0^{\dagger} \text{div} Y^{\mathrm{C}} 
= s^{\mathrm{A}} + s^{\mathrm{B}} + s^{\mathrm{C}}.
$
This linearity allows for a detailed decomposition of the total potential into the contributions of its subgroups.
In contrast, some traditional measures, e.g., in-migration to out-migration ratio, do not satisfy this equality.

Next, our measure of empirical liveability --- the potential --- exhibits spatial autocorrelation.
In our case, using a fixed distance band of 10km to define neighbourhood relationships, Moran's $I$ index was 0.12 ($p< 0.001$) for FwSC and 0.27 ($p < 0.001$) for WoRA.
This is an expected feature, as the liveability of a location is determined not only by its intrinsic factors but also by those of surrounding areas.
For example, large-scale retailers attract consumers from surrounding areas, and we accounted for this spillover effect in our regression model.
Nevertheless, the spatial dependence can still lead to inefficient and biased estimates.
Therefore, to more accurately estimate the direct effects of various factors and policy implementations, the regression analysis presented here could be enhanced by employing spatial models or causal inference methods with panel data.

In addition, we provide a note on the spatial unit of analysis. Although we selected the level of a municipality as the unit of analysis in this paper, the proposed method is applicable to any non-overlapping spatial units, such as provinces, metropolitan areas, or even grid cells (Unlike the cases in the USA and the UK, the Japanese Statistics Bureau does not offer an official delineation of the metropolis. Alternatively, researchers defined the Metropolitan Employment Areas \autocite{Kanemoto2002UEA}).

\section{Methods}
\subsection{Municipalities in Japan}
\label{sec:municipalities}
Japan's local governance is organized into a two-tier system:
The upper tier consists of 47 prefectures, and the lower tier comprises the municipalities.
The municipalities include cities, towns and villages.
According to a Japanese law, a city must generally have
(1) a population of at least 50,000 residents,
(2) at least 60 percent of households in a central urbanised area,
(3) at least 60 percent of the population engaged in urban occupations,
and (4) any other urban infrastructure or conditions specified by the prefectural ordinance.

\subsection{Regional utility from migration data as potentials}
\label{sec:utility} 
A related but different approach to the problem we addressed is to estimate regional utility in an economic model of migration decisions from migration data.
First, we investigate the formulas of utility, calculated in related studies in economics literature.
Next, as our contribution to these works, we provide an explicit formula of utility using the Hodge decomposition, which has been implicitly given as coefficients of a regression equation.
We reinterpret the utilities from the viewpoint of the combinatorial Hodge theory.

In Douglas \& Wall \autocite{douglas2000measuring}, Wall\autocite{wall2001voting} and Nakajima \& Tabuchi \autocite{nakajima2011estimating}, utility was numerically determined as coefficients in the regression analysis (More precisely, the discussion here applies to a version of regression equation of Douglas \& Wall \autocite{douglas2000measuring} and Nakajima \& Tabuchi \autocite{nakajima2011estimating} which includes regional dummies only).
For example, Nakajima \& Tabuchi \autocite{nakajima2011estimating} formulated the following Ordinary Least Square (OLS) regression equation:
\begin{align}
  \ln \frac{M_{ij}}{M_{ji}} &=  \sum_{k=2}^N b_k D_k + e_{ij}, \quad i,j = 1, \cdots, N \, \text{and} \, i \neq j
  \label{eq:nakajima_regression}
\end{align}
where $M$ is the OD matrix of migration flow, $N$ is the number of locations, and $b_k$ is the regression coefficient proportional to the utility: $b_k = u_k/\alpha$. $D_k$ is a dummy variable given by
\begin{align*}
  D_k = \begin{cases}
    1 & \quad (k =j)\\
    -1 &\quad (k =i)\\
    0 &\quad \text{otherwise}
  \end{cases}.
\end{align*}
$e_{ij}$ is the residual term.
This regression is based on the formula between utility $u$ and migration flow $M$,
which is derived from a discrete choice model under approximations:
\begin{align*}
  \ln \frac{M_{ij}}{M_{ji}}  = \frac{2}{\alpha} (u_j - u_i).
\end{align*}

A series of studies by Douglas \& Wall \autocite{douglas2000measuring} derived another formula between $u$ and $M$ under different approximations:
\begin{align}
  \frac{M_{ij} - M_{ji}}{P_i P_j}  = \kappa (u_j - u_i), \label{eq:douglas1997}
\end{align}
where $P_i, P_j$ are the populations at locations $i$ and $j$, respectively, and $\kappa$ is a constant.

The OLS equation \eqref{eq:nakajima_regression} corresponds to the following optimisation problem:
\begin{align*}
  \min_{u}  \sum_{i}^N \sum_{j \neq i}^N \left[ \frac{2}{\alpha}(u_j - u_i) - \ln \frac{M_{ij}}{M_{ji}} \right]^2. %
\end{align*}
This optimisation problem is solved by projecting on subspace im(grad).
Here we consider another edge flow $Y_{ij} = \ln \frac{M_{ij}}{M_{ji}}$,
instead of net migration $M_{ij} - M_{ji}$,
and then, $u_i$ is explicitly obtained as a potential function in the combinatorial Hodge theory:
\begin{align}
 u_i = 
 \frac{\alpha}{2 N}\sum_{j} \left[ \ln \frac{M_{ji}}{M_{ij}} \right] = 
 \frac{\alpha}{2 N} \ln \left[ \prod_j \frac{M_{ji}}{M_{ij}} \right]. \label{eq:nakajima_utility}
\end{align}

This derived expression provides a clear interpretation of the utility.
The utility $u$ is a monotonically increasing function of the product of the incoming-and-outgoing flow ratio $M_{ji}/M_{ij}$. 
The utility is undefined if a pair of zones $(i,j)$ has no flow $M_{ij} = 0$.
This issue is problematic for the higher spatial resolution of migration datasets because the number of migrants can be small or zero.
Nakajima \& Tabuchi \autocite{nakajima2011estimating} analysed the migrations by a coarse-grained region level or the prefecture level.
However, at the municipal level, we found no migration between many pairs of municipalities.

Similarly, we obtained the explicit equation from formula \eqref{eq:douglas1997},
\begin{align}
 u_i = 
 \frac{1}{\kappa N}\sum_{j \neq i}^N \left[ \frac{M_{ji} - M_{ij}}{P_i P_j} \right]
 =
 \frac{1}{\kappa N P_i} \left[ \sum_{j \neq i}^N \frac{M_{ji}}{P_j} - \sum_{j \neq i}^N \frac{M_{ij}}{P_j}  \right]. \label{eq:douglas_utility}
\end{align}
This expression indicates that the incoming and outgoing per-capita flow, $M_{ji}/P_j$ and $M_{ij}/P_j$, are aggregated over all other locations, and the balance between their sums determines the utility $u_i$.

Notably, in the above OLS equations and their corresponding optimisation problems, all pairs of locations are included without weighting.
In this condition, distance factors are neglected, as mentioned in the case of net migration. %

As another contribution of this analysis, 
the utilities estimated in the previous works are guaranteed to be of an interval scale, not just an ordinal scale.
Moreover, the metric is globally consistent, as shown in Figure \ref{fig:illustration}; the accumulated differences via a path from one location to another are always equal to those via another path between the same endpoints.
Thus, the utilities provide consistent global rankings.
In other words, using the framework of the combinatorial Hodge theory, we can find the closest point in the desirable subspace to the given non-transitive migration data.

\subsection{Distance between municipalities}
\label{sec:method_distance}
We calculated the distance from municipality $i$ to municipality $j$ as follows:
The origin point in municipality $i$ and the destination point in municipality $j$ are stochastically selected in proportion to the population census.
Over the ensemble of this random selection, we defined the road distance $d_{ij}$ between municipalities as the mean distance:
$$
d_{ij} = \sum_{u \in S_i} \sum_{v \in S_j} \frac{P_u}{\bar{P}_i} \frac{P_v}{\bar{P}_j} d_{u \to v},
$$
where $S_i$ and $S_j$ are sets of census blocks (Japanese MESH3 Boundaries) within municipalities $i$ and $j$, respectively. $P_u$ and $P_v$ are the populations in the census blocks $u$ and $v$,
and $\bar{P}_i$ and $\bar{P}_j$ are the total populations in municipalities $i$ and $j$.
$d_{u \to v}$  is the road distance from the centroid of block $u$ to the centroid of block $v$, calculated using the Open Source Routing Machine (OSRM) based on the Open Street Map \autocite{OSRM}.

We excluded municipalities for which distance calculation is inaccurate from our sample.
Because of the limitation of the Open Street Map dataset, it has been reported that this distance calculation could provide an inaccurate distance when marine transportation is required to travel from an origin to a destination \autocite{Tonda2022OSRM}.
We excluded municipalities in Okinawa and other small islands connected mainly by ocean ferries and airplanes.
After the exclusions, 1813 municipalities were included in the target area.

\section{Data availability}
Migration datasets and the census data that support the findings of this study are publicly available from Official Statistics Bureau of Japan portal site (\url{https://www.e-stat.go.jp/}).

\section{Code availability}

Scripts used in this study are publicly available via Github at \url{https://github.com/TakaakiAokiWork/HodgePotentialHumanFlow}.

\printbibliography

\clearpage
\appendix

\renewcommand{\thefigure}{\arabic{figure}}
\renewcommand{\thetable}{\arabic{table}}
\renewcommand{\theequation}{\arabic{equation}}

\section{Datasets of regression analysis}
In this section, we describe additional information on datasets used in the regression analysis. %
The following datasets are publicly available from the Official Statistics Bureau of Japan (\url{https://www.e-stat.go.jp/}), except for the crime dataset, which is publicly available from the portal site of the Japanese national police agency (\url{https://www.npa.go.jp/toukei/seianki/hanzaiopendatalink.html}). 

In this study, we analysed migrations in 2019 and used the data from 2019 as explanatory variables, unless otherwise described.

\textit{Population} is the number of residents in each municipality, based on the Japanese basic resident registers, including all nationalities.

\textit{Habitable Area} is a habitable area of each municipality, described in `Statistical reports on land area by prefecture and municipality in Japan.'
The habitable area is defined as the total land area minus the forest and lake areas.

\textit{Income} is based on `Survey of municipal taxation status of Japan.'
This is calculated as the aggregate income of each municipality divided by the number of taxpayers.

\textit{Land price} is based on `Publication of market value of standard sites by prefectural government', available for almost all municipalities.
We selected the sites for residential areas and evaluated the mean value for each municipality.

\textit{Labour Force Participation Rate (LFPR)} and \textit{Unemployment rate} are based on `Population census of Japan', which is reported every five years.
We used a 2020 dataset.
In the dataset, LFPR is defined as the percentage of the labour force over the population aged over 15 years, and the unemployment rate is defined as the percentage of unemployed persons in the labour force.

\textit{Manufacturing} and \textit{Business Service} are the percentage of manufacturing and business service employees over the total employees, respectively.
The data are based on the 2016 `Economic Census for Business Activity of Japan.'

\textit{Human capital} is the percentage of graduated persons over the 25-aged population,
based on the 2020 `Population census of Japan.'

\textit{Houses far from preschool} is the percentage of houses whose nearest preschool is over one kilometre apart. The data is based on the 2018 `Housing and Land Survey of Japan.'

\textit{ST Ratio Elementary} is the student--teacher ratio in elementary schools, based on the `School Basic Survey of Japan.'

\textit{Library} is the number of libraries, based on the 2018 `Social education Survey of Japan.'

\textit{Large Scale Retailers 10km} counts the numbers of large-scale retailers in each municipality and its neighbouring municipalities, based on the 2016 `Economic Census for Business Activity of Japan.'
In the census, retailers with more than 50 employees are categorised as large-scale retailers.

\textit{Mean commuting time} is calculated by the commuting time of the households whose primary income is obtained by a commuter, as described in the 2018 `Housing and Land Survey of Japan.' 
The commuting time was categorised into four groups: 0--30 min, 30--60 min, 60--90 min, and over 90 min. We set the representative times of these groups as 15, 45, 75, and 115 min, respectively, and obtained the mean commuting time weighted by the number of households in each group.

\textit{Housing starts} is the number of houses under construction.
\textit{Vacancy rate} is the percentage of unoccupied houses over total houses.
\textit{Owner occupied} is the percentage of owner-occupied houses over the total.
These data are based on the 2018 `Housing and Land Survey of Japan.' 
\textit{Sewage tank} is the percentage of residents with proper sewage disposal facilities.
The data is obtained from the `Nation Survey on the State of Discharge and Treatment of Municipal Solid Waste.'

\textit{Hospital per capita} is the number of hospitals per capita, reported in `Survey of Medical Institutions.'

\textit{Thefts per capita} is the number of thefts per capita.
According to a report by the National Police Agency in Japan ({\url{https://www.npa.go.jp/publications/statistics/index.html}), theft accounts for over 70 percent of crimes.
``Japanese crime open datasets' provide specific theft type that mainly occur out of doors.
We counted the number of thefts that occurred outdoors and used these as crime indicators in public spaces. 

\textit{City class} becomes one only if a municipality is the core city or `government ordinance city' on the first day in 2019.
The core and government ordinance cities are listed on the Ministry of Internal Affairs and Communications website in Japan ({\url{https://www.soumu.go.jp/}).

The dummy variables starting with \textit{Region} are generated from a categorical variable of Japanese regions based on the definition of the Statistics Bureau of Japan ({\url{https://www.stat.go.jp/data/shugyou/1997/3-1.html}), setting the baseline to the \textit{Shikoku} region.

\section{Descriptive statistics of explanatory variables in regression analysis}
Supplementary Table \ref{tab:regression_descriptive} summarises the descriptive statistics of the explanatory variables used in the regression analysis. %

\begin{table}[!h]
\centering
\caption{\label{tab:regression_descriptive}Descriptive statistics}
\centering
\fontsize{8}{10}\selectfont
\begin{tabular}[t]{llllllll}
\toprule
Variable & N & Mean & Std. Dev. & Min & Pctl. 25 & Pctl. 75 & Max\\
\midrule
Population & 1813 & 0.0691 & 0.0993 & 0.000369 & 0.00963 & 0.0859 & 0.917\\
HabitableArea & 1813 & 0.661 & 0.7 & 0.015 & 0.204 & 0.852 & 8.05\\
Income & 1813 & 2.97 & 0.611 & 2.08 & 2.6 & 3.2 & 12.2\\
LandPrice & 1811 & 5.62 & 12.5 & 0.152 & 1.08 & 5.05 & 291\\
LFPR & 1813 & 60.4 & 5.11 & 0 & 57.6 & 63.4 & 90\\
UnemploymentRate & 1813 & 3.68 & 1.09 & 0 & 3.08 & 4.22 & 10.6\\
Manufacturing & 1812 & 20.6 & 12.3 & 0 & 11 & 28.1 & 69.9\\
BusinessService & 1812 & 5.56 & 4.41 & 0 & 3.04 & 6.69 & 58.8\\
HumanCapital & 1812 & 15.8 & 7.5 & 3.44 & 10.3 & 19.5 & 47.2\\
HousesFarFromPreschool & 1215 & 39.7 & 24.6 & 0 & 19.3 & 58.6 & 100\\
STRatio\_Elementary & 1805 & 12.2 & 4.3 & 0.667 & 9.07 & 15.6 & 22.1\\
Library & 1813 & 1.81 & 2.34 & 0 & 1 & 2 & 26\\
LargeScaleRetailers\_10km & 1813 & 0.691 & 1.19 & 0.000015 & 0.0472 & 0.691 & 8.04\\
MeanCommutingTime & 1215 & 0.525 & 0.154 & 0.27 & 0.413 & 0.606 & 1.05\\
HousingStarts & 1215 & 0.0681 & 0.172 & 0 & 0.01 & 0.07 & 4.18\\
VacancyRate & 1215 & 14.6 & 5.69 & 3.39 & 10.9 & 17.2 & 68.2\\
OwnerOccupied & 1215 & 56.6 & 15 & 8.49 & 46.6 & 67.2 & 96.4\\
SewageTank & 1813 & 0.891 & 0.133 & 0 & 0.842 & 0.986 & 1\\
HospitalPerCaptia & 1813 & 0.0729 & 0.0747 & 0 & 0.0236 & 0.0994 & 0.795\\
TheftPerCaptia & 1813 & 1.17 & 1.25 & 0 & 0.353 & 1.59 & 16.5\\
CityClass & 1811 & 0.109 & 0.312 & 0 & 0 & 0 & 1\\
\bottomrule
\end{tabular}
\end{table}
 
\section{Additional regression analysis}
Supplementary Table \ref{tbl:regression_robust} presents the results of the linear regression analysis when several non-significant explanatory variables were removed.

\begin{table}[!htbp] \centering 
  \caption{Linear regression analysis: robustness check} 
  \label{tbl:regression_robust} 
\scriptsize 
\begin{tabular}{@{\extracolsep{1pt}}lD{.}{.}{-3} D{.}{.}{-3} D{.}{.}{-3} D{.}{.}{-3} } 
\\[-1.8ex]\hline 
\hline \\[-1.8ex] 
 & \multicolumn{4}{c}{\textit{Dependent variable:}} \\ 
\cline{2-5} 
\\[-1.8ex] & \multicolumn{2}{c}{Women of reproductive age} & \multicolumn{2}{c}{Families with small children} \\ 
\hline \\[-1.8ex] 
 Population & -0.035 & -0.007 & -2.083^{***} & -2.169^{***} \\ 
  HabitableArea & -0.336^{***} & -0.327^{***} & -0.029 & -0.031 \\ 
  Income & -0.052 & -0.055 & -0.220^{*} & -0.207^{*} \\ 
  LandPrice & -0.006 & -0.006 & 0.011^{*} & 0.008 \\ 
  LFPR & 0.080^{***} & 0.083^{***} & -0.002 & 0.005 \\ 
  UnemploymentRate & 0.038 & 0.042 & 0.051 & 0.071 \\ 
  Manufacturing & -0.006 &  & 0.0001 &  \\ 
  BusinessService & -0.004 &  & -0.012 &  \\ 
  HumanCapital & 0.056^{***} & 0.056^{***} & 0.042^{***} & 0.040^{***} \\ 
  HousesFarFromPreschool & 0.005^{*} & 0.005^{*} & -0.002 & -0.002 \\ 
  STRatio\_Elementary & 0.061^{***} & 0.059^{***} & 0.083^{***} & 0.092^{***} \\ 
  Library & 0.011 & 0.010 & 0.037^{*} & 0.037^{**} \\ 
  LargeScaleRetailers\_10km & 0.343^{***} & 0.348^{***} & -0.009 & 0.005 \\ 
  MeanCommutingTime & -2.324^{***} & -2.007^{***} & 1.847^{***} & 1.997^{***} \\ 
  HousingStarts & 0.221 & 0.240 & 0.571^{**} & 0.579^{**} \\ 
  VacancyRate & 0.004 & 0.007 & -0.001 & -0.005 \\ 
  OwnerOccupied & 0.011^{*} & 0.010^{*} & 0.058^{***} & 0.058^{***} \\ 
  SewageTank & -0.255 &  & 0.693 &  \\ 
  HospitalPerCaptia & -1.319 &  & -1.410 &  \\ 
  TheftPerCaptia & 0.012 &  & 0.034 &  \\ 
  CityClass & 0.021 &  & -0.044 &  \\ 
  RegionChugoku & -0.138 & -0.118 & -0.175 & -0.146 \\ 
  RegionHokkaido & 0.609^{**} & 0.650^{***} & 1.206^{***} & 1.199^{***} \\ 
  RegionHokyriku & -0.029 & -0.014 & -0.203 & -0.142 \\ 
  RegionKinki & -0.128 & -0.123 & -0.692^{***} & -0.604^{***} \\ 
  RegionKyusyu & 0.220 & 0.227 & 0.417^{**} & 0.351^{*} \\ 
  RegionNorthKantoKoshin & -0.180 & -0.179 & -0.250 & -0.198 \\ 
  RegionSouthKanto & 0.013 & 0.004 & -0.480^{*} & -0.444^{*} \\ 
  RegionTohoku & 0.143 & 0.212 & 0.010 & -0.008 \\ 
  RegionTokai & -0.227 & -0.222 & -0.518^{**} & -0.427^{**} \\ 
  Constant & -6.160^{***} & -6.907^{***} & -6.590^{***} & -6.791^{***} \\ 
 \hline \\[-1.8ex] 
Observations & \multicolumn{1}{c}{1,212} & \multicolumn{1}{c}{1,212} & \multicolumn{1}{c}{1,212} & \multicolumn{1}{c}{1,212} \\ 
R$^{2}$ & \multicolumn{1}{c}{0.450} & \multicolumn{1}{c}{0.447} & \multicolumn{1}{c}{0.357} & \multicolumn{1}{c}{0.352} \\ 
Adjusted R$^{2}$ & \multicolumn{1}{c}{0.436} & \multicolumn{1}{c}{0.435} & \multicolumn{1}{c}{0.341} & \multicolumn{1}{c}{0.339} \\ 
\hline 
\hline \\[-1.8ex] 
\textit{Note:}  & \multicolumn{4}{r}{$^{*}$p$<$0.05; $^{**}$p$<$0.01; $^{***}$p$<$0.001} \\ 
\end{tabular} 
\end{table} 
 
\section{Large images of the panels in Figures 3 and 5}
For better visibility, larger versions of the panels showing the map of Japan in Figures 3 and 5,
are shown in this section.

\begin{figure}
    \centering
    \includegraphics[width=\linewidth]{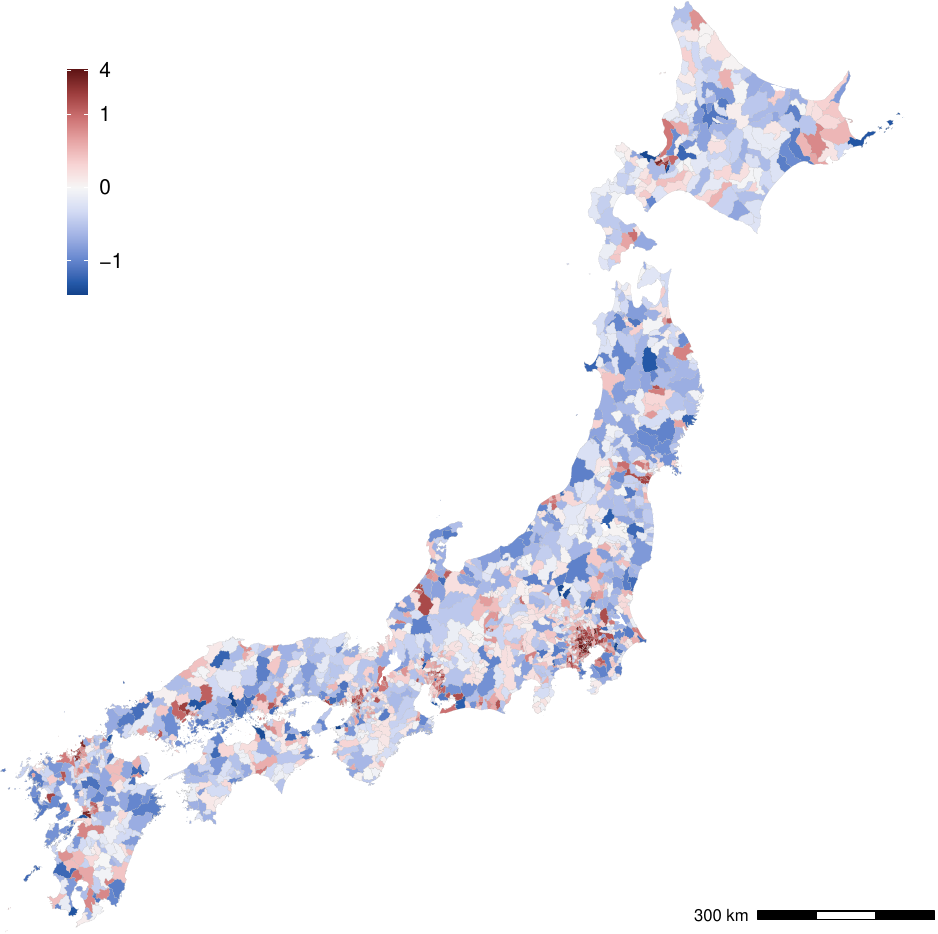}
  \caption{
    A larger image of Figure \ref{fig:potential}A,
    showing the potential landscape of migration flows for \textit{women of reproductive age} in the entire target area in Japan.
  } 
\end{figure}

\begin{figure}
    \centering
    \includegraphics[width=\linewidth]{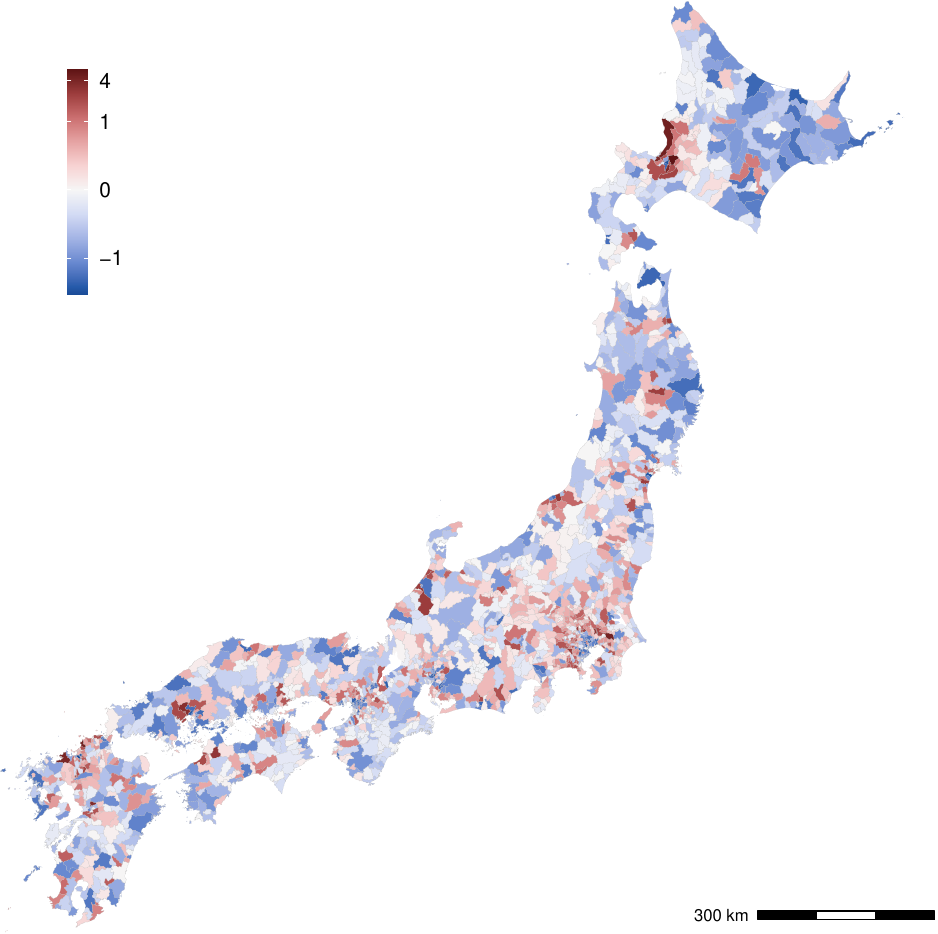}
  \caption{
    A larger image of Figure \ref{fig:potential}C,
    showing the potential landscape of migration flows for \textit{families with small children} in the entire target area in Japan.
  } 
\end{figure}

\begin{figure}
    \centering
    \includegraphics[width = \linewidth]{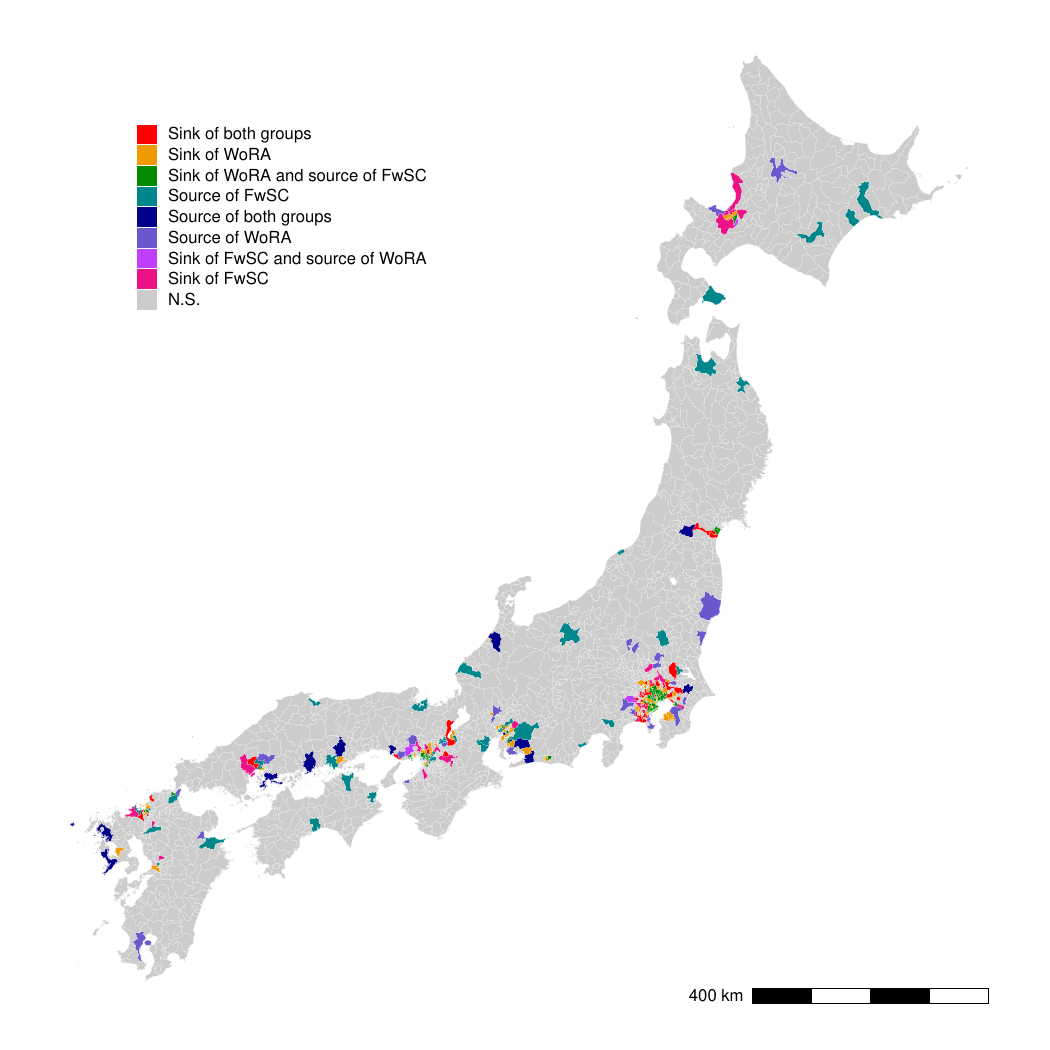}
    \caption{
      A larger image of Figure \ref{fig:classification}B,
      showing the map of the classified municipalities in the entire target area in Japan.
Significant municipalities of migrations are indicated by colours, according to the classification scheme shown in Figure \ref{fig:classification}A.
}
\end{figure}

\end{document}